\documentclass[aps,preprint,superscriptaddress]{revtex4}
\usepackage{graphicx}
\usepackage{color}
\usepackage{bm}

\begin{document}
published in: Phys. Rev. A 81, 043426 (2010)

\title
{
Auger decay of $1\sigma_g$ and $1\sigma_u$ hole states of N$_2$ molecule:
disentangling decay routes from coincidence measurements
}

\author{S.~K.~Semenov}
\affiliation{State University of Aerospace Instrumentation, 190000,
St. Petersburg, Russia}
\author{M.~S.~Sch{\"o}ffler}

\author{J.~Titze}
\author{N.~Petridis}
\author{T.~Jahnke}
\author{K.~Cole}
\author{L.~Ph.~H.~Schmidt}
\author{A.~Czasch}

\affiliation{Institut f{\"u}r Kernphysik, University Frankfurt, 
Max-von-Laue-Str. 1, D-60438 Frankfurt Germany}

\author{D.~Akoury}
\affiliation
{Institut f{\"u}r Kernphysik, University Frankfurt, 
Max-von-Laue-Str. 1, D-60438 Frankfurt Germany}
\affiliation
{Lawrence Berkeley National Lab., Berkeley CA 94720}

\author{O.~Jagutzki}
\affiliation
{Institut f{\"u}r Kernphysik, University Frankfurt, 
Max-von-Laue-Str. 1, D-60438 Frankfurt Germany}

\author{J.~B.~Williams}
\affiliation
{Department of Physics, Auburn University Auburn AL 36849}

\author{T.~Osipov}
\author{S.~Lee}
\author{M.~H.~Prior}
\author{A.~Belkacem}
\affiliation
{Lawrence Berkeley National Lab., Berkeley CA 94720}

\author{A.~L.~Landers}
\affiliation
{Department of Physics, Auburn University Auburn AL 36849}

\author{H.~Schmidt-B{\"o}cking}
\affiliation
{Institut f{\"u}r Kernphysik, University Frankfurt, 
Max-von-Laue-Str. 1, D-60438 Frankfurt Germany}

\author{Th.~Weber}
\affiliation
{Lawrence Berkeley National Lab., Berkeley CA 94720}

\author{N.~A.~Cherepkov}
\affiliation{State University of Aerospace Instrumentation, 190000,
St. Petersburg, Russia}
\affiliation{Institut f{\"u}r Kernphysik, University Frankfurt, 
Max-von-Laue-Str. 1, D-60438 Frankfurt Germany}

\author{R.~D{\"o}rner}
\affiliation
{Institut f{\"u}r Kernphysik, University Frankfurt, 
Max-von-Laue-Str. 1, D-60438 Frankfurt Germany}

\begin{abstract}
Results  of  the  most sophisticated measurements  in
coincidence  of  the angular resolved $K$-shell  photo-
and  Auger-electrons, and of two atomic ions produced
by   dissociation  of  N$_2$  molecule,  are analyzed.
Detection of photoelectrons at certain angles  allows
separating the Auger decay processes of the  $1\sigma_g$ and 
$1\sigma_u$ core  hole  states. The Auger  electron  angular
distributions  for  each of  these  hole  states  are
measured as a function of the kinetic energy  release
of   two  atomic  ions  and  are  compared  with  the
corresponding   theoretical  angular   distributions.
From   that   comparison  one  can  disentangle   the
contributions  of different repulsive doubly  charged
molecular  ion  states to the Auger decay.  Different
kinetic  energy  release values are directly  related
to  the different internuclear distances. In this way
one  can  trace  experimentally the behavior  of  the
potential  energy curves of dicationic  final  states
inside  the Frank-Condon region. Presentation of  the
Auger  electron angular distributions as  a  function
of kinetic energy release of two atomic ions opens  a
new dimension in the study of Auger decay.
\end{abstract}

\maketitle

\section{Introduction}
      Auger  decay studies of molecules have a long history,
though it seems that not all characteristics of that process
have  been investigated up to now. In atoms the Auger  decay
corresponds  to  a  transition between  two  (quasi)discrete
states,  therefore  main  attention  in  the  atomic   Auger
electron   spectroscopy   studies   was   focused   on   the
identification  of  discrete lines  [1,2].  The  coincidence
study  of photoelectrons and Auger electrons enabled a  much
more detailed study of complex Auger electron spectra with a
high  precision [3-8]. As compared to atoms, in diatomic 
molecules due to lower symmetry (axial instead of spherical 
in the case of atoms) two new degrees of freedom appear, the 
vibrational and the rotational motion of nuclei. Due  to
that  the molecular Auger electron spectra are substantially
modified.  The rotational splitting is too small to be 
resolved in the Auger electron spectra, while  the vibrational 
splitting is of the same  order of  magnitude as the Auger 
line widths and can broaden them. In addition to  that,  the
photoionization  followed by Auger decay produces  a  doubly
charged  molecular  ion  which  often  dissociates  creating
atomic  ions in their ground or excited states. In that  way
the  excitation  energy of the initial state is  distributed
among the Auger electron, the nuclear motion and electronic
excitation of the final products. Instead of a well  defined
discrete  line a broad continuum of Auger electron  energies
appears.  This continuum cannot be identified by its  energy
position  since usually several transitions are contributing
at  the  same  energy. Therefore a basically new  method  is
needed  for studying the continuous Auger electron  emission
spectra in  molecules. Such a method must take  advantage
of  the  axial symmetry of diatomic molecules  in  order  to
extract  additional  information not available  in  the standard
Auger electron spectra. Namely, when the dissociation process
is faster than the rotational motion, the latter can be 
disregarded which opens the possibility to study the Auger 
decay of fixed-in-space molecules. That is the way to get
the most detailed information about the Auger decay.

     Let  us consider the photoionization of the $K$-shell  of
N$_2$  molecule which  produces highly excited  molecular ion
state.  Within a short time of about 7 fs this state decays,
predominantly  by emission of a fast Auger electron  (around
360  eV).  As  a result, a doubly charged molecular  ion  is
created  with two holes in valence shells. At the next  step
this  doubly charged molecular ion dissociates predominantly
into  two  N$^{+}$ atomic ions with the kinetic  energy release
(KER) in the region of 4 to 20 eV. The dissociation time  is
usually  short compared to the molecular rotation, therefore
the  direction  of  motion  of the  atomic  ions  gives  the
direction   of   the  molecular  axis   at   the   time   of
photoabsorption and Auger decay.

  Earlier  the  Auger decay of core ionized N$_2$  molecules
has been studied by different methods, in particular, by the
Auger electron spectroscopy [9-10], and KER spectroscopy  of
the two N$^{+}$ ions [11-13]. In these studies, as in the case 
of atomic  Auger  decay, mainly the resonance  structures  have
been  investigated. We report on the most detailed study  of
the  Auger  decay  process by detecting in  coincidence  the
photoelectron, the Auger electron and the two atomic  singly
charged ions (all of them being energy and angular resolved)
using  the  COLd  Target  Recoil Ion  Momentum  Spectroscopy
(COLTRIMS) technique [14-15]. The single hole states in  the
$K$-shell  of  N$_2$  molecule are due to  symmetry  requirements
split into two states, $1\sigma_g$ and $1\sigma_u$, and it is of interest 
to separate  the  Auger decay processes of  these  two  states.
Their energy splitting is rather small, about 100 meV, which
is  nearly equal to the width of these states equal  to  120
meV.  Nevertheless, recent very high resolution measurements
allowed  resolving these states in the photoelectron spectra
[16,17]  as  well as in the Auger electron spectra  [18,19].
For  the  Auger decay routes leading to dissociative  states
the  symmetry of  Auger electron cannot be deduced from  its
continuous  energy  alone, but the  initial  singly  ionized
state has to be determined. That could in principal be  done
by  measuring the Auger electron energy and the KER  with  a
resolution of better than the $g/u$ splitting. Alternative one
can  measure the photoelectron in coincidence to  the  Auger
electron and deduce the character of the $K$-hole from  either
the  photoelectron energy or the emission angle. Since the  
necessary
energy  resolution  is  hard to  achieve  in  a  coincidence
experiment, we have opted to use the photoelectron angle  to
tag the g or u core hole state.

   The angular distributions of photoelectrons from the $K$-shell  
of N$_2$ molecule have been studied theoretically, and from 
calculations  it is known  that at  some angles predominantly 
$1\sigma_g$ or $1\sigma_u$ shell is contributing  [14-15].  By
measuring   the  Auger  electron  angular  distribution   in
coincidence  with  the  photoelectrons  collected  at  these
angles one can separate the contributions of $1\sigma_g$ 
and $1\sigma_u$
shells  to  the Auger decay process without need to  resolve
these  transitions in energy [15]. As is  shown  below,  the
corresponding  Auger  electron  angular  distributions  for
transitions from the $1\sigma_g$ and $1\sigma_u$ hole states 
strongly depend 
on the configuration and  the term of the  final  dicationic
state. Comparing experimental and theoretical Auger electron
angular distributions, one can identify the transitions into
different  dicationic states. Important is that this  method
allows  studying  mainly the continuous part  of  the  Auger
spectrum which is hard to study by any other method [20,21].
This continuum is formed by Auger transitions into repulsive
doubly charged molecular ion states which do not create  any
resonance  structure.  However, the  most  intense  resonant
Auger  transitions can also be studied by this  method.  The
preliminary  results of this study have  been  published  in
[22].

\section{Theory}
      A  detailed  description of the  method  used  in  our
calculations has been presented earlier in [23-25].  Here we
shall  mention  mainly the modifications introduced  in  the
present  calculations. We describe theoretically the angular
distributions of photoelectrons from core levels  and  Auger
electrons  measured in coincidence with each other,  and  in
coincidence  with the atomic ions resulted from dissociation
of  doubly charged molecular ion. The dissociation  time  is
implied  to  be  much shorter than the period  of  molecular
rotation,  so  that the direction of motion of  dissociation
products  gives the direction of molecular axis at the  time
of  the  photoionization  and the  Auger  decay.  Since  the
dissociation  step  is  not  considered  theoretically,   we
calculate the photoionization and the Auger decay of  fixed-
in-space  molecules.  We  imply that  a  two-step  model  is
applicable according to which the photon absorption is  much
faster  than  the Auger decay [1,2]. Under these  conditions
the  amplitude of the process can be presented as a  product
of a dipole $d$ and a Coulomb $V$ matrix elements
\begin{equation}\label{eq:1}
f_{f,i}^\lambda  ({\vec p}_A ,{\vec p}) = 
\left\langle {\Psi _f^{N - 2} 
\psi _{{\vec p}_A }^{-}  \left| V \right|\Psi _i^{N - 1} } 
\right\rangle \left\langle {\Psi _i^{N - 1} 
\psi _{\vec p}^-  \left| {d_\lambda  } \right|\Psi _0 } 
\right\rangle .
\end{equation}
Here $\left| {\Psi _0 } \right\rangle$  means the ground state 
wave function of  a  molecule  containing $N$ electrons, 
$\lambda$ is projection of a photon  angular  momentum  in  
a photon frame with the  $z$  axis directed  along
the  photon  beam, $\Psi _i^{N - 1}$  and $\Psi _f^{N - 2}$
 are a singly charged and  a  doubly
charged molecular ion wave functions of the intermediate and
final states, respectively, $\psi_{\vec p}^- $
 and $\psi _{{\vec p}_A }^- $ are the photoelectron  and
the  Auger  electron wave functions defined in the molecular
frame, and $\vec p$ and $\vec p_A$  are the moments of the
photoelectron and the Auger electron, respectively.  In  our
case  the  intermediate state is the  state  with  one  hole
either  in the $1\sigma_g$ or in the $1\sigma_u$ shell. The 
final state $\Psi _f^{N - 2}$ has two holes in the valence 
shells.

  Doubly  differential cross section for the process of core 
ionization of N$_2$ molecule with a subsequent Auger decay
in which both photo- and Auger-electrons are ejected at some
fixed  angles  is  given within the two-step  model  by  the
equation
\begin{equation}\label{eq:2}
\frac{{d\sigma _{fi}^\lambda  }}
{{d\Omega _{p_A } d\Omega _p }} \propto 
\left| {f_{fi}^\lambda  ({\vec p}_A ,{\vec p})} \right|^2 ,\,\,\,i 
= 1\sigma _g \,or\,\,1\sigma _u .
\end{equation}
  Since the Lorentzian widths of the $1\sigma_g$ and $1\sigma_u$ 
photoelectron lines in  N$_2$ are approximately  equal  to their  
energy splitting,  in the photoelectron-Auger electron coincidence
experiment  the photoelectrons from these shells  cannot  be
energetically resolved [15]. This situation is described theoretically by
treating  the  $1\sigma_g$ and $1\sigma_u$ states as if they were 
degenerate. Then instead of (2) we get
\begin{equation}\label{eq:3}
\frac{{d\sigma _f }}
{{d\Omega _{p_A } d\Omega _p }} \propto 
\left| {f_{f,1\sigma _g } ({\vec p}_A ,{\vec p}) + f_{f,1\sigma _u } 
({\vec p}_A ,{\vec p})} \right|^2 .
\end{equation}
Now  we  have a square modulus of the sum of two  amplitudes
which includes also the interference term, and this equation
actually describes a deviation from the two-step model.  The
role  of the interference term in equation (3) was discussed
in [15,26].

       In   the  present  analysis  we  selected  only   the
photoelectron  ejection  angles  at  which  the  predominant
contribution is given by one of two $K$-shells, that is  where
one  of the following conditions is fulfilled,
$f_{f,1\sigma _g } ({\vec p}_A ,{\vec p}) 
<< f_{f,1\sigma _u }({\vec p}_A ,{\vec p}),$
or  
$f_{f,1\sigma _g } ({\vec p}_A ,{\vec p}) 
>> f_{f,1\sigma _u }({\vec p}_A ,{\vec p}),$
In  these cases the interference  term is small  and  to  a 
good approximation  can  be  neglected, so that only a square
modulus of one of the two transitions in equation (3)  gives
a  substantial contribution. Then all the general  equations
presented in [25] are valid here, too. 

    Our   calculations  have  been  performed  in   prolate
spheroidal coordinates by the method described in [23].  The
two  steps  (the  photoionization and the Auger  decay)  are
treated  in the following way. At first the single  electron
wave  functions of the ground state of the neutral  molecule
are calculated in the Hartree-Fock (HF) approximation. After
that  the wave functions for the intermediate singly charged
molecular  ion state are calculated in the relaxed  core  HF
(RCHF)  approximation as a solution of the HF equation  with
the   potential  formed  by  the  self-consistent  HF   wave
functions  of  a  singly  charged  ion.  The  relaxed   core
approximation  allows taking into account the  rearrangement
of  the  molecular orbitals to the creation of a  core  hole
state.  But  the  usual integer charge 1 for  the  ion  core
overestimates  this  effect,  therefore  we   proposed   the
modification  of  this method by using a fractional  charge.
The  latter  is selected empirically from the  condition  to
correctly  describe the position in energy of the $\sigma^*$ 
shape resonance of the photoionization cross section. For  the 
$K$-shell  of N$_2$ the best agreement with experiment  was found
with   the   fractional  charge  equal  to  0.7  [24].   The
photoelectron wave function is calculated in the RCHF  field
and  is  orthogonalized to the ground state wave  functions.
With  the  wave functions described above the dipole  matrix
elements are calculated according to the equations (10)-(11)
of  [25].  Many-electron correlations in the photoionization
process   are  taken  into  account  in  the  random   phase
approximation by solving the corresponding equation for  the
dipole matrix elements presented in [23]. 
 
    The initial state for the Auger decay is described  by
the  same self-consistent RCHF wave functions, 
$\varphi _j^{(i)} $  of the  singly
charged  molecular ion  as in the photoionization step.  For
the doubly charged final molecular ion state another set  of
the  self-consistent HF wave functions 
$\varphi _j^{(f)} $  is calculated,  this
time  with  the  integer charge 2. The Auger  electron  wave
function  is  calculated in the frozen field of  the  doubly
charged  ion.  The Auger decay amplitude is defined  by  the
Coulomb matrix element given by equations (34)-(40) of [25].
Since  the  wave  functions $\varphi _j^{(i)} $ 
and  $\varphi _j^{(f)}$ are  not  orthogonal,  
we calculate also the overlap matrix between the HF orbitals 
of the initial and final states 
${\mathbf{S}}_{jk} = \left\langle {\varphi _j^{(f)} 
\left| {\varphi _k^{(i)} } \right.} \right\rangle $
and obtain the Auger amplitude
following the procedure proposed in [27]. The Auger electron
energy  in  the particular cases considered here  is  large,
about  360  eV,  so  that the contribution of  many-electron
correlations   is  expected  to  be  small.   Therefore   we
restricted calculations by the HF approximation  as  it  was
done already in [25] for CO molecule.

\section{Experiment}
      The  experiment was performed at beamline 11.0.2 of the
Advanced  Light  Source of Lawrence Berkeley Laboratory  via
COLTRIMS  technique [28-31]. A supersonic gas  jet,  with  a
precooled  nozzle  provided  an  internally  cold  and  well
localized target of N$_2$ molecules in their vibrational ground
state.  This gas jet was intersected by a beam of circularly
polarized   photons   (419  eV)  from   beamline 11.0.2. The
interception volume of well below 0.3 mm$^3$ was situated in a
region  of  homogeneous  parallel  electric  (12  V/cm)  and
magnetic  (6.5 G) fields. The fields were prependicular  to
the  gas jet. The fields guided the photoelectrons toward  a
multichannel plate detector (diameter 80 mm) with delay-line
position  readout  [32]. The fields  assured $4\pi$ collection
solid  angle  for the photoelectrons, while the  fast  Auger
electrons  where  detected only within the  small  geometrical
solid  angle. They were used for calibration purposes  only.
In case the N$_2^{2+}$ ions break up, the resulted ionic
fragments gain a large amount of kinetic energy from the
Coulomb explosion. Therefore the solid angle of detection
depends on the orientation of the molecular axis at the
instant of fragmentation: those  N$_2^{2+}$ ions that 
fragmented within $15^\circ$ parallel to the electric field 
axis of our spectrometer were guided toward a second   
position-sensitive  detector,  72   cm   from   the
interaction point. From the position of impact and the  time
of  flight of the photoelectron and ions, we could determine
their   vector   momenta, respectively. To improve the ion 
momentum resolution, we used a three-dimensional time and 
space-focusing ion optics setup (see figure 12 in [28]). Momentum
vectors of the photoelectron and the two ions from the four-
body  final  state  were  measured  directly,  whereas   the
momentum  of  the fourth particle, the Auger  electron,  was
obtained  through momentum conservation. This  was  possible
only  because  the lens system avoided the deterioration  of
the ion momentum resolution due to the spatial extension  of
the interaction volume and since the N$_2$ jet was sufficiently
cold in the direction of the gas beam due to cooling of  the
nozzle.  For the nozzle conditions great care was  taken  to
avoid  clustering of the beam while maintaining  its  narrow
momentum spread.

     The  experiment  yielded  the  full $4\pi$ solid  angle
distribution of the Auger electron and photoelectron and  1\%
solid  angle  for the ion momentum. We obtained  an  overall
resolution of better than 50 meV (see fig 8 below)  for  the
KER and 0.5 atomic unit momentum resolution of the center of
mass  motion (i.e. the momentum of the Auger electron). The data  
were recorded in list mode, so any combination of angles and
energies of the particles could be sorted out in the off-line
analysis without repeating the experiment. The dataset  used
in  the  present  analysis is the same as  in  [15,22].  All
spectra  reported were taken simultaneously  with  the  same
apparatus to reduce possible systematic errors.

\section{Results}
\subsection{The basis of the method}
     Fig.  1a shows the theoretical angular distribution  of
photoelectrons in the molecular frame ejected  from  $K$-shell
of N$_2$ molecule  by circularly polarized  light  at  photon
energy 419 eV. This energy corresponds to the well known $\sigma^*$
shape resonance in the photoabsorption cross section [33] so
that the photoelectron intensity at this photon energy has a
maximum.  The  calculations have been  performed  with  many
electron  correlations  taken  into  account  in   the   RPA
approximation. It is seen that at the angles 60-80 and  240-
260 degrees  the predominant contribution is given  by  the
photoelectrons ejected from $1\sigma_g$ shell. At the angles 
$140^\circ-150^\circ$ and  $320^\circ-330^\circ$,  vice versa, 
the  predominant contribution is given by the $1\sigma_u$ shell. 
Therefore to a  good approximation  one  can  say that  by  
measuring  the  Auger electron  angular  distribution  in  
coincidence  with   the photoelectrons collected at the angles 
mentioned  above  one can study the Auger decay process 
separately for the $1\sigma_g$ and $1\sigma_u$  shells  
without  need to resolve these  transitions  in
energy  [15].  In  the  experiment the  photoelectrons  were
collected  from  a  broader range  of  angles  in  order  to
increase the intensity of the signal. Namely, the angles 
$35^\circ-85^\circ$ and $215-265^\circ$ were used to select the 
contribution of $1\sigma_g$ shell,  and  the  angles  
$115-150^\circ$ and  $295-330^\circ$  for the contribution  
of  the  $1\sigma_u$ shell. As  is  evident  from  the
figures, the separation of the contributions of the $1\sigma_g$  
and $1\sigma_u$  shells is not complete, there is always some  
admixture of the state of the opposite parity which must be taken 
into account while comparing theory with experiment (see below).
\begin{figure}
 \includegraphics[width=6.5cm,angle=0]{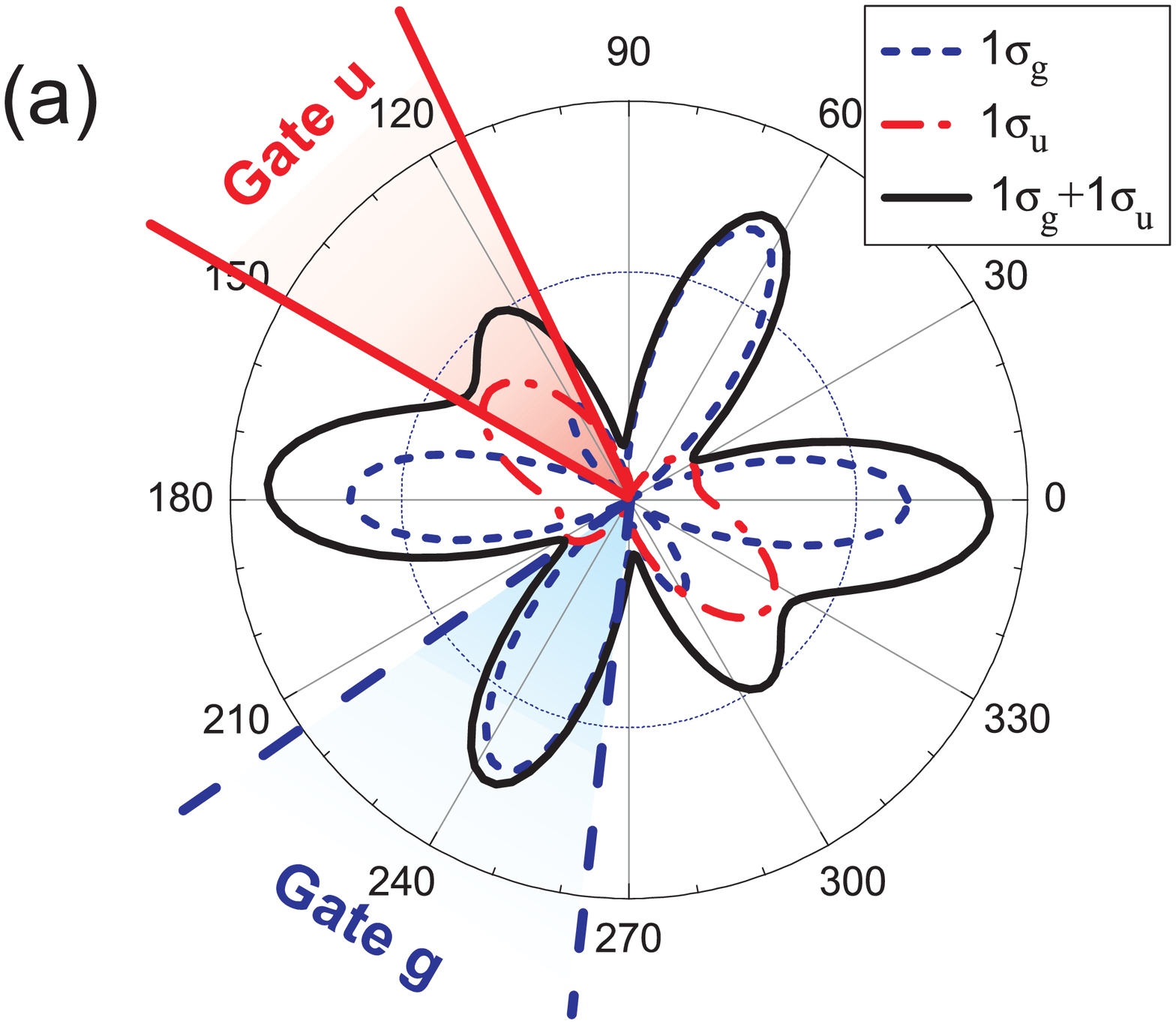}
 \includegraphics[width=7.5cm,angle=0]{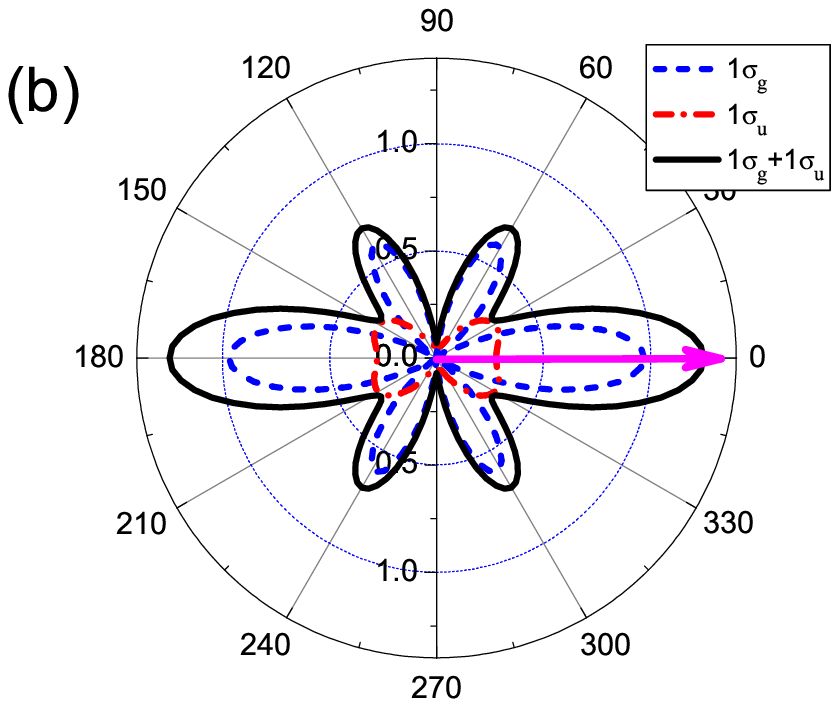}
 \includegraphics[width=7.0cm,angle=0]{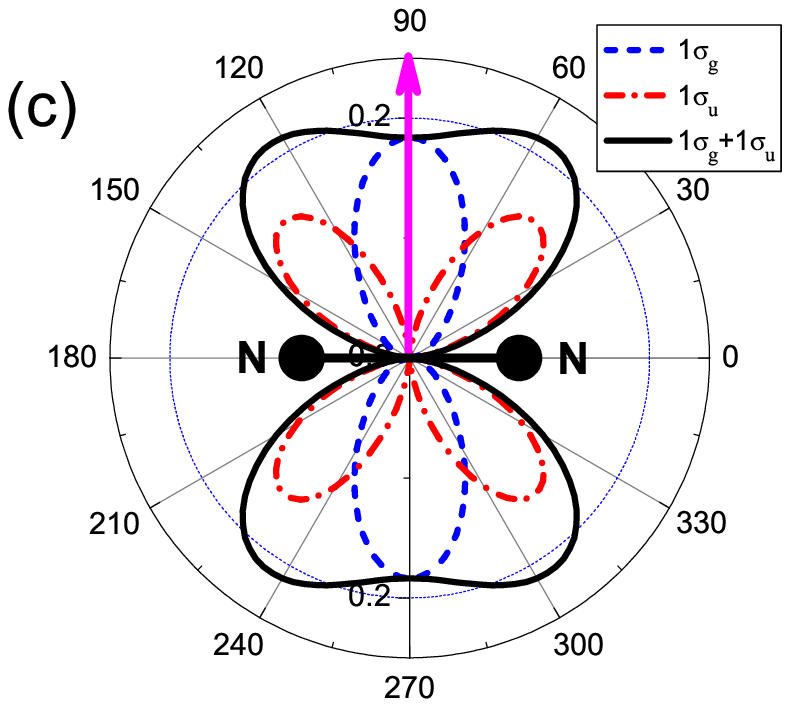}
 \includegraphics[width=7.5cm,angle=0]{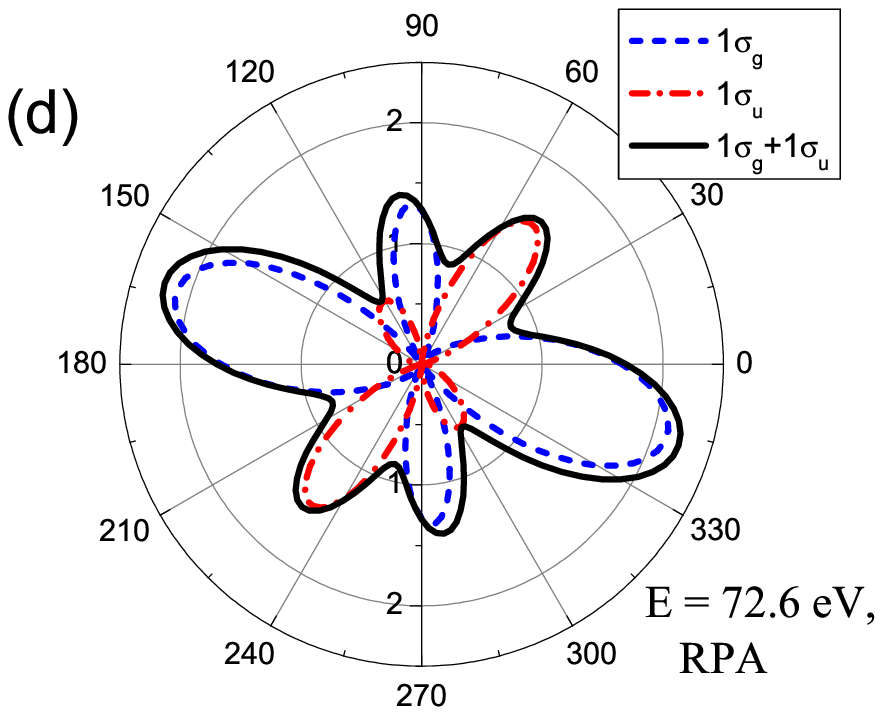}
\caption
{ (Color  online)
 Molecular frame photoelectron angular distributions
in the plane perpendicular to the photon beam
calculated for several light polarizations and photon 
energies: (a) for left  handed circularly
polarized  light  at photon energy 419 eV; (b) for 
light linearly polarized parallel to the molecular axis
at 419 eV; (c) for light linearly polarized perpendicular 
to the molecular axis at 419 eV; (d) for left circular
polarization at photon energy 483 eV. Molecular  axis  is
directed along the horizontal axis as is shown in (c). 
The contributions of 
$1\sigma_g$ and $1\sigma_u$ hole  states  are shown by 
dashed  and  dot-dashed lines, respectively. Their sum is 
shown by solid line.
}
\end{figure}

     It  is  worth  while  to  mention  that  absorption  of
circularly  polarized  light gives a better  opportunity  to
separate  the  contributions of the $1\sigma_g$ and $1\sigma_u$
shells  as compared  to linearly polarized light. Figs. 1b,c  show  
the photoelectron angular distributions for absorption of  light
linearly  polarized  parallel and perpendicular to the molecular  
axis, respectively, for the same photon energy 419 eV.  In  both
cases  one can easily separate the contribution of the $1\sigma_g$
shell, while the contribution of the $1\sigma_u$ shell is greatly
overlapping  with the $1\sigma_g$ one and hardly can be separated.
The  other question is whether the particular photon  energy
can be favorable or unfavorable for such a separation, or an
experiment  can  be performed at any photon  energy.  As  an
example,  we  show  in  Fig.  1d the  photoelectron  angular
distribution for circularly polarized light at photon energy
483  eV (the photoelectron energy 73 eV). Here one can  also
quite well separate  the contributions of the $1\sigma_g$ and $1\sigma_u$  
shells.  So, the  method can be applied at different photon 
energies  and is not bound to the shape resonance.

     To  interpret  the Auger electron spectra corresponding
to  the  decay  of  the $1\sigma_g$ or $1\sigma_u$ hole state  
we performed calculations of Auger electron angular distributions 
for all possible final doubly charged molecular ion states with  
two holes  in the outermost $3\sigma_g$, $1\pi_u$, or $2\sigma_u$  
shells.  Single configuration  approximation was used in these 
calculations. These   angular  distributions strongly depend on 
the configuration  and the term of the dicationic  final  state.
Since in the experiment mainly the  dissociating  states  are
contributing,  we concentrate  on the consideration of these
states. Fig. 2 shows the potential energy curves for several
states  of the N$_2^{2+}$ ion taken from the references [11, 34-36].
As  we  already mentioned, the photoionization and the Auger
decay processes are fast compared to the nuclear motion, and
the  internuclear  distance during these  processes  remains
weakly changed. Therefore only the Franck-Condon (FC) region
is contributing to the formation of doubly charged molecular
ions.  The  vertical axis in Fig. 2 gives the KER energy  of
two  N$^+$ ions  after  dissociation  process.  The  part  of
potential energy curves inside the FC region gives the range
of   KER  energies  to  which  the  corresponding  term   is
contributing.
\begin{figure}
 \includegraphics[width=7.5cm,angle=0]{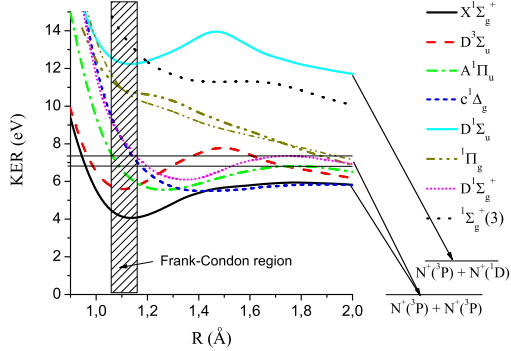}
\caption
{ (Color  online)
 Potential energy curves from refs. [11,34-36]  for
several final dicationic states mentioned in the figure. The
zero  KER corresponds to the dissociation limit  into
the N$^+(3P)$ + N$^+(3P)$ ion states. For the $^1\Pi_g$ final 
state two curves  are shown, one from ref. [35] (bold curve), 
and  the other from ref. [36] (thin curve). The horizonthal
lines mark the positions of potential energy barriers for the
A$^1\Pi_u$ and D$^1\Sigma_g^+$ terms.
}
\end{figure}
\begin{figure}
 \includegraphics[width=7.5cm,angle=0]{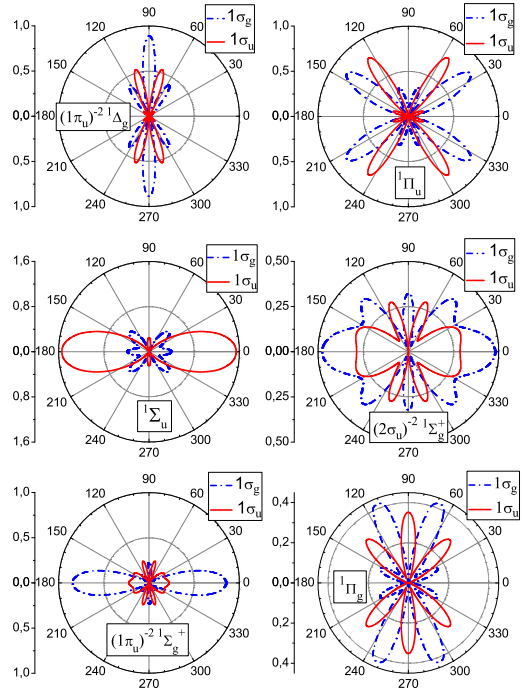}
\caption
{ (Color  online)
 Theoretical Auger electron angular distributions for
several  final  dicationic states mentioned in  the  figures
giving   the   main  contribution  to  the  Auger   electron
intensity. The molecular axis is directed along the horizontal axis.
}
\end{figure}

       Theoretical  Auger electron angular distributions for several
final  dicationic states giving the predominant contribution
to the Auger electron intensity are presented in Fig. 3. Our
calculations  are  in agreement with the earlier  result  of
{\AA}gren  [37] according to which the triplet final states  are
giving  rather  small contribution to the  Auger  decay  and
hardly  can be disentangled in our experiment, therefore  we
do not show them. There is one exception, the $^3\Sigma_u$ term 
(see Fig.  2),  which  has  a local minimum  and  for  which the
quasidiscrete  final states are observed and  identified  as
discussed  below.  Similarly there are quasidiscrete  states
corresponding to the $^1\Sigma_u$ term which has a local  minimum  in
the potential energy curve, too.
\begin{figure}
 \includegraphics[width=7.5cm,angle=0]{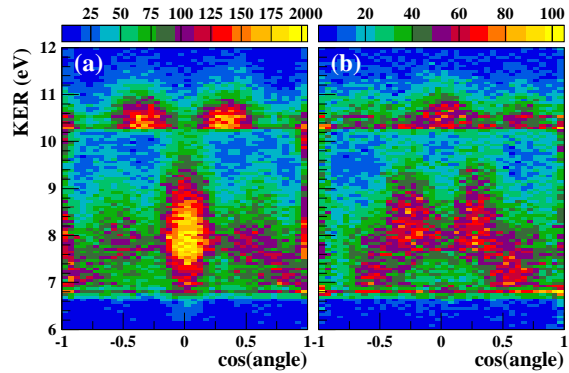}
\caption
{ (Color  online)
 Experimental Auger electron intensities measured  as
a function of cosine of the ejection angle $\theta$ relative to 
the molecular axis (horizontal axis) and of KER (vertical axis).
(a)  and  (b) correspond to the Auger decay of $1\sigma_g$  
and  $1\sigma_u$ hole states, respectively.
}
\end{figure}

 Now  we have enough information to start the analysis of the
experimental results. Fig. 4 shows the angular distributions
of the Auger electrons for a photon energy 419 eV measured in
coincidence   with  the  photoelectrons   ejected   at   the
directions corresponding to ionization of either $1\sigma_g$ or 
$1\sigma_u$ shell in accord with Fig. 1a. In this way the Auger  
decay processes of the $1\sigma_g$ and $1\sigma_u$ states are 
separated. They are shown  separately  in  Figs. 4a and  4b,  
respectively. The vertical  axis  in these figures corresponds  
to the KER energy.  As a function of KER one  can  single  out  
three regions corresponding to KER energies 7-7.5 eV, 7.5-9.5  eV,
and  10.3-11.5  eV,  where  the angular  distributions  have
different characteristic features. Comparing theoretical and
experimental  angular distributions one  can  determine  the
main  Auger  decay  channels contributing  at  a  given  KER.
Since the KER for any final state is defined
by  the internuclear distance at which the Auger decay takes
place,  the  analysis  of the KER dependence  of  the  Auger
electron   angular  distributions  allows  determining   the
internuclear distances at which a given Auger decay  channel
contributes. The separation of the Auger decay processes of 
the $1\sigma_g$ and $1\sigma_u$ core holes plays the key 
role in this analysis.

\subsection{Analysis   of  the  coincidence  Auger   
electron-photoelectron spectra}
   Let  us start from the KER energies 7-7.5 eV. From Fig. 2  
follows  that  three final states are contributing  here,
$(3\sigma_g)^{-1}(1\pi_u^{-1})$ $^1\Pi_u$, $(1\pi_u)^{-2}$ 
$^1\Sigma_g^+$, and $(1\pi_u)^{-2}$ $^1\Delta_g$.  Fig. 5
shows  the  comparison  of  theoretical  results  with   the
experimental data (in arbitrary units). Since in theory  the
dissociation  process  is  not considered,  the  theoretical
angular  distributions are not connected with  any  definite
value of KER, while in experiment we have a contribution  of
a  well  defined KER energy region. Therefore  the  relative
contributions of the three final states mentioned above  are
taken  theoretically as free parameters fitted by comparison
with  the  experiment. The result of this fitting gives  for
the  relative  contributions of these states  the  following
ratio: 
$I(^1\Pi_u) : I(^1\Sigma_g^+) : I(^1\Delta_g) = 1 : 0.7 : 0.7$. 
Fig. 5a and 5b show the  relative  contributions of these  
transitions  together with  their sum. For the $1\sigma_g$ hole 
state (Fig. 5a)  the  main maximum at $90^\circ$ is given by 
the $^1\Delta_g$ term, the maxima at about $30^\circ$ and 
$150^\circ$ are due to the $^1\Pi_u$ term, and finally the only
nonzero  contribution at $0^\circ$ and $180^\circ$ is given by  
the $^1\Sigma_g^+$ term. For the $1\sigma_u$ hole state (Fig. 5b) 
the main maxima at $55^\circ$ and $125^\circ$ are defined by the 
$^1\Pi_u$ term, while the two maxima at $75^\circ$ and $105^\circ$ 
due to the $^1\Delta_g$ term are making the main maxima broader.
The  only contribution at $0^\circ$ and $180^\circ$ is given 
again by the $^1\Sigma_g^+$ term. In Figs. 5c and 5d the total 
contributions  from  figs. 5a  and 5b are shown by dashed curves. 
The agreement
between  theory and experiment is satisfactory. Let us  take
now   into   account  the  fact  that  the   separation   of
contributions  of the $1\sigma_g$ and $1\sigma_u$ shells in the  
coincidence experiment is not complete as is evident from Fig. 1a.  
We must  allow  some  admixture of the hole state  of  the  opposite
parity  to each angular distribution. With such an admixture
(added with a fitted parameter) the theoretical curves shown
by solid lines in Fig. 5c and 5d are coming to a fairly good
agreement  with  the experiment. The main  lobes  and  their
relative intensities are correctly reproduced by the theory.
\begin{figure}
 \includegraphics[width=7.5cm,angle=0]{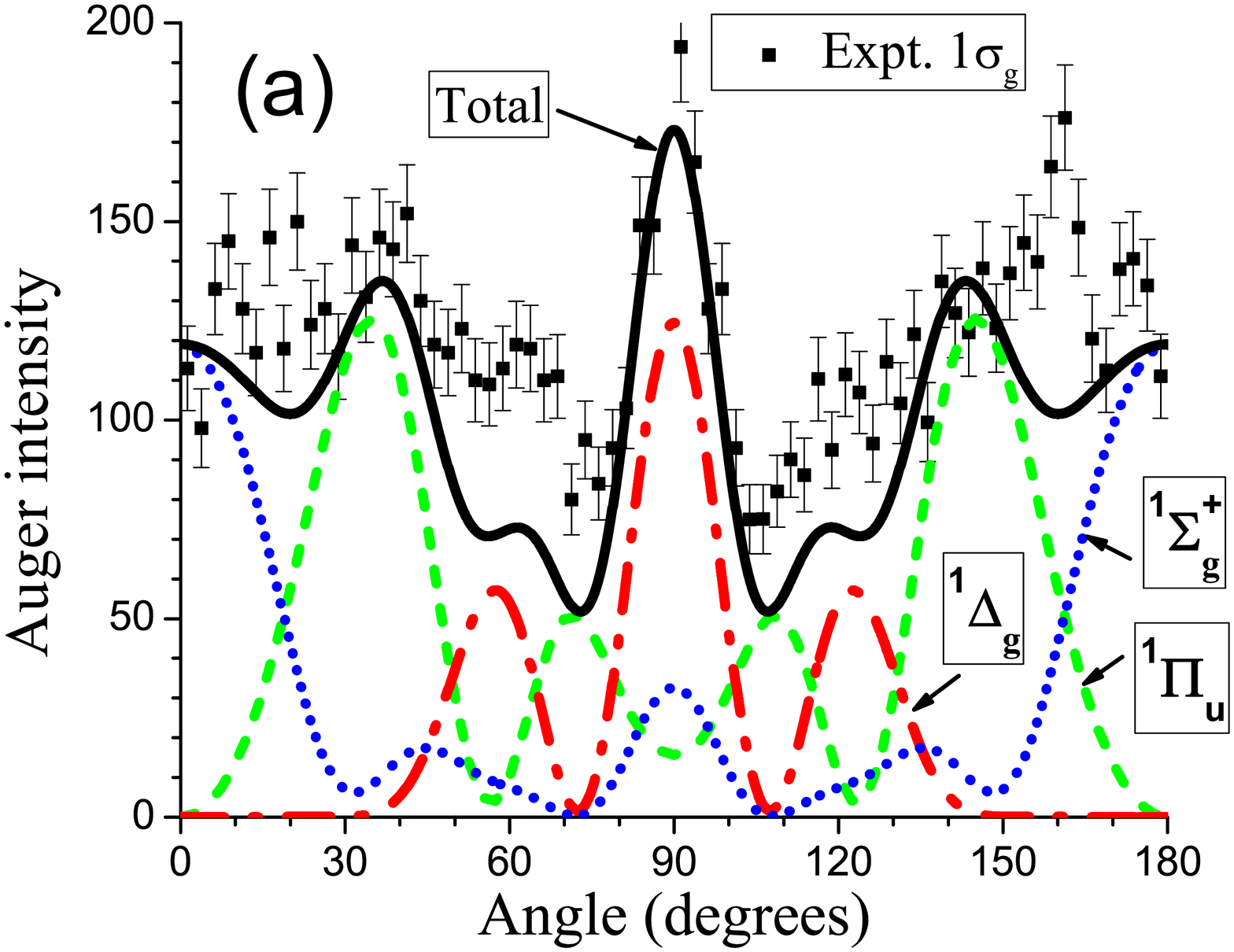}
 \includegraphics[width=7.5cm,angle=0]{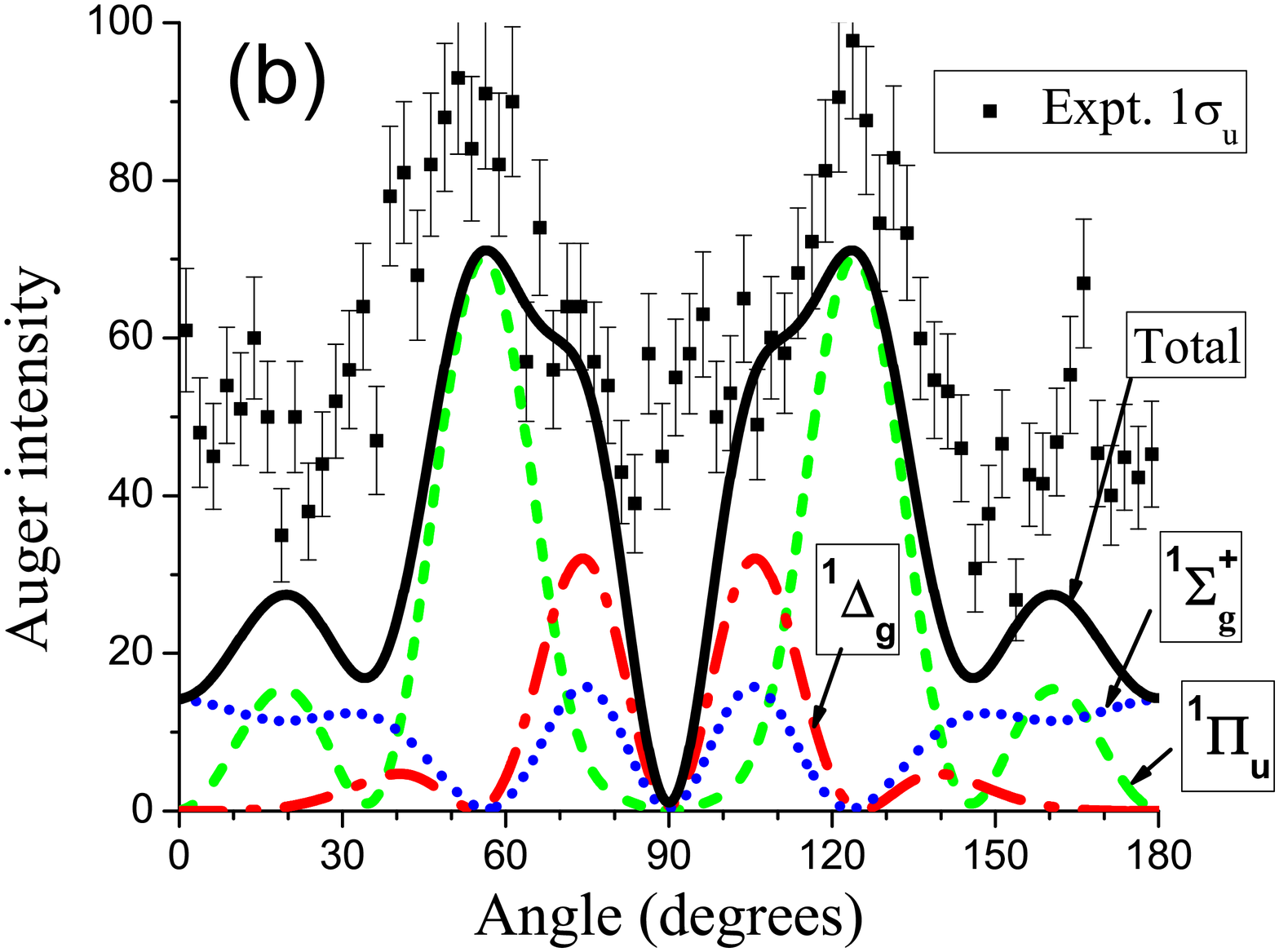}
 \includegraphics[width=7.5cm,angle=0]{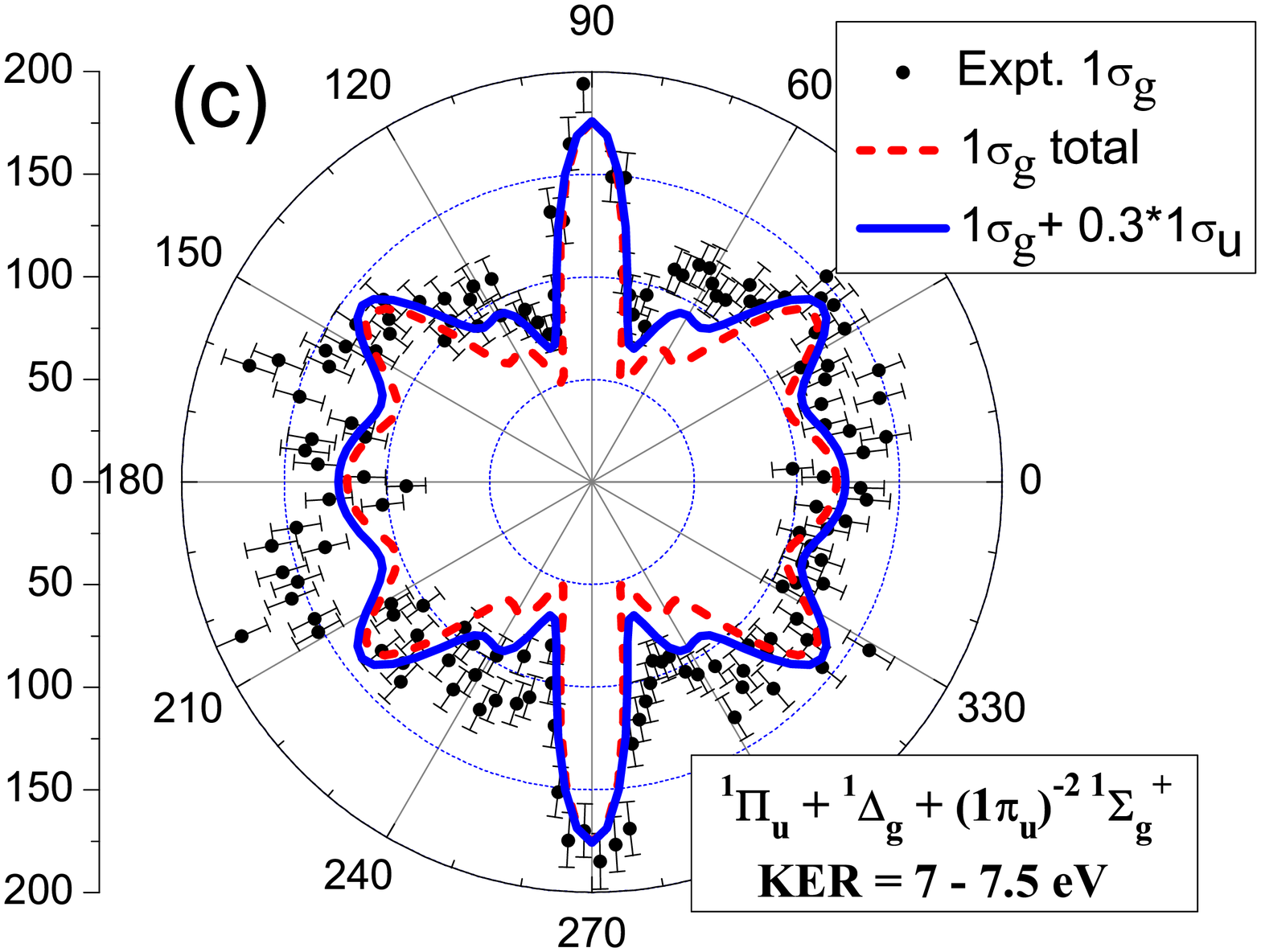}
 \includegraphics[width=7.5cm,angle=0]{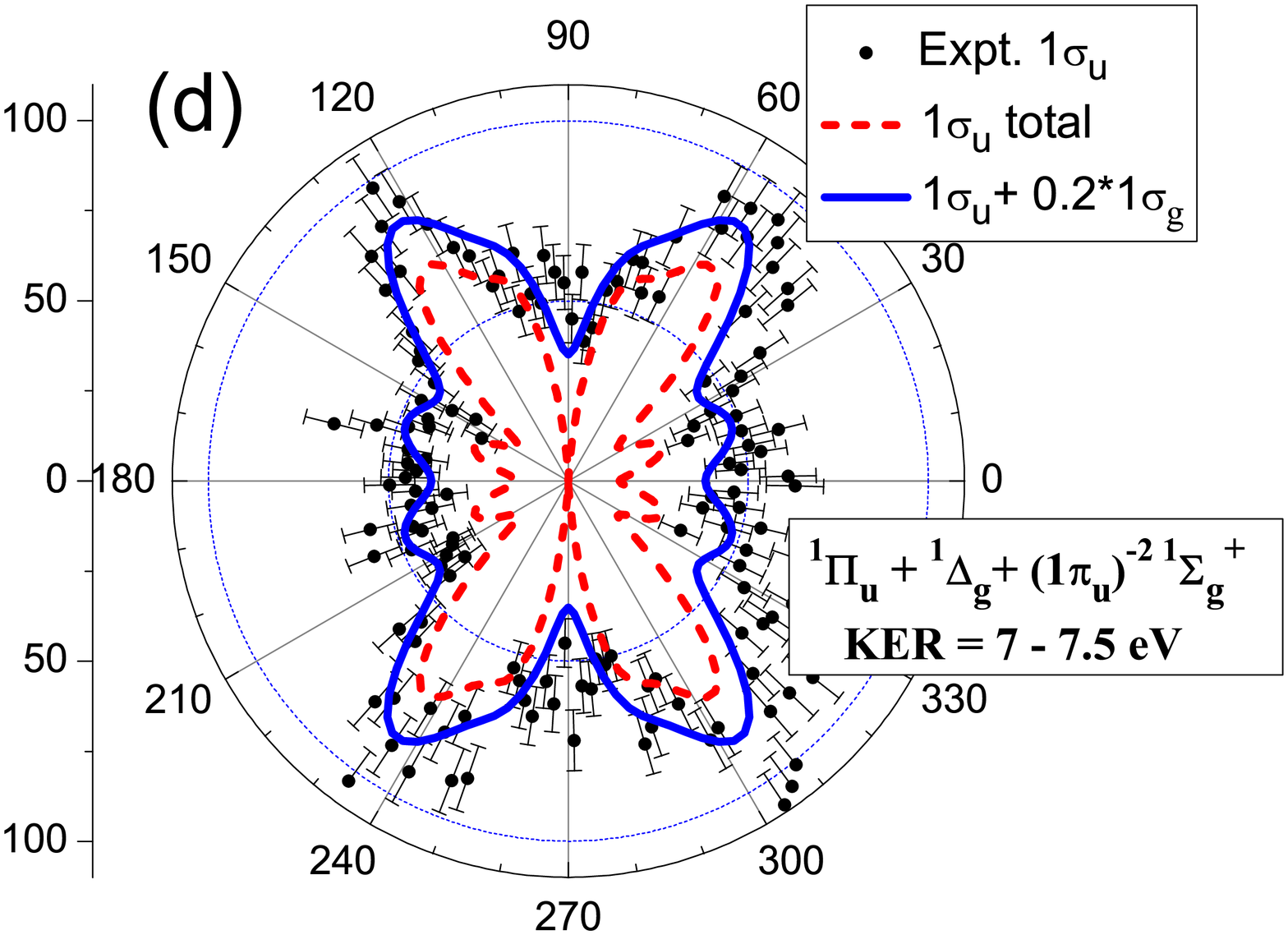}
\caption
{ (Color  online)
 Auger electron angular distributions  (in arbitrary units) measured  
in coincidence  with  photoelectrons (points with  error  bars)
corresponding  to ionization of either $1\sigma_g$ (a,c)  
or  $1\sigma_u$  (b,d) shell,  integrated  over 
KER energies  from  7  to  7.5  eV.
Molecular  axis is directed along the horizontal  axis.  
The theoretical calculations (normalized to  the
experiment) include the Auger transitions 
to  the following doubly charged  molecular  ion states 
$(3\sigma_g)^{-1} (1\pi_u)^{-1}$ $^1\Pi_u$, 
$(1\pi_u)^{-2}$ $^1\Delta_g$, and 
$(1\pi_u)^{-2}$ $^1\Sigma_g^+$ . The
dashed  and solid lines in (c), (d)  show
the  results of calculation without and with the inclusion
of the admixture of the hole state of the opposite parity 
correspondingly (see the text for  detail).
}
\end{figure}

     Fig. 6 shows the comparison of calculated and measured
Auger electron angular distributions at KER energies from  8
to  9  eV.  In  accord with Fig. 2 only two  doubly  charged
molecular  ion  states $(1\pi_u)^{-2}$ $^1\Delta_g$ and  
$(1\pi_u)^{-2}$ $^1\Sigma_g^+$ are contributing here. 
The $(1\pi_u)^{-2}$ $^1\Delta_g$ state is 
responsible  (i) for  the intensive lobe at the ejection angle 
$90^\circ$ (above the horizontal  axis) and two smaller lobes 
at $57.5^\circ$ and  $122.5^\circ$ for  the $1\sigma_g$ state, 
and (ii) for the intensive lobes at  the angles  $75^\circ$  
and $105^\circ$ for the $1\sigma_u$ state. The $(1\pi_u)^{-2}$  
$^1\Sigma_g^+$ state  contributes mainly along the molecular  
axis  at  the angles  $0^\circ$  and  $180^\circ$ (qualitatively 
similar results  though without  resolving the contributions 
of  $1\sigma_g$ and $1\sigma_u$  hole states  have  been 
obtained theoretically in  [20]).  Dashed lines  again show 
the results obtained for pure $1\sigma_g$ or  pure $1\sigma_u$
hole  states. The relative contributions  of  different
terms  in the fitted curve is 
$I(^1\Delta_g) : I(^1\Sigma_g^+) = 1:0.87$. For the $1\sigma_u$ 
state agreement with experiment is only qualitative. But
after  adding the contribution of the state of the  opposite
parity  shown  by solid lines in Fig. 6 the  agreement  with
experiment  is  becoming quite satisfactory. The  amount  of
admixture is defined by fitting to the experiment.
\begin{figure}
 \includegraphics[width=7.5cm,angle=0]{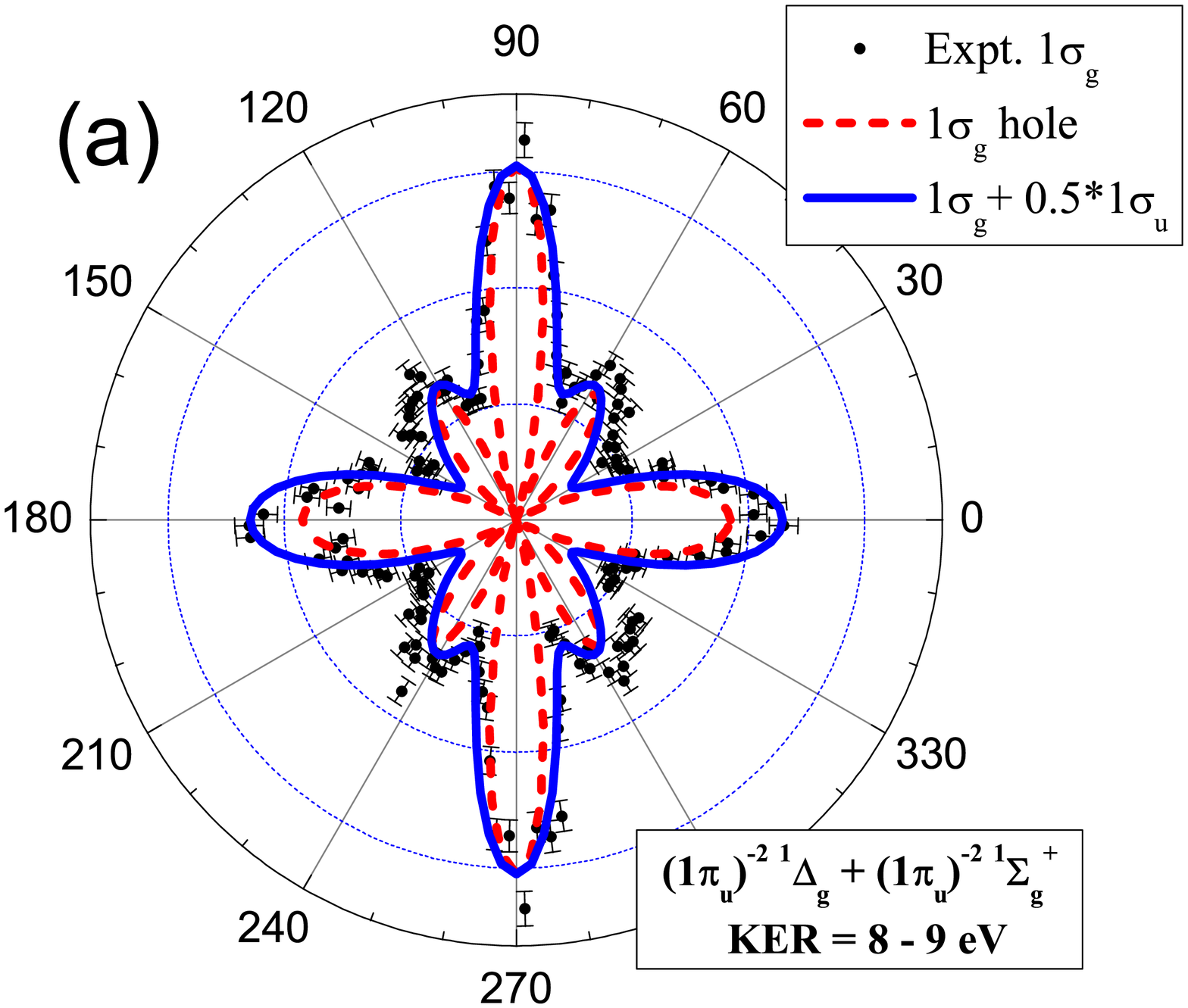}
 \includegraphics[width=7.5cm,angle=0]{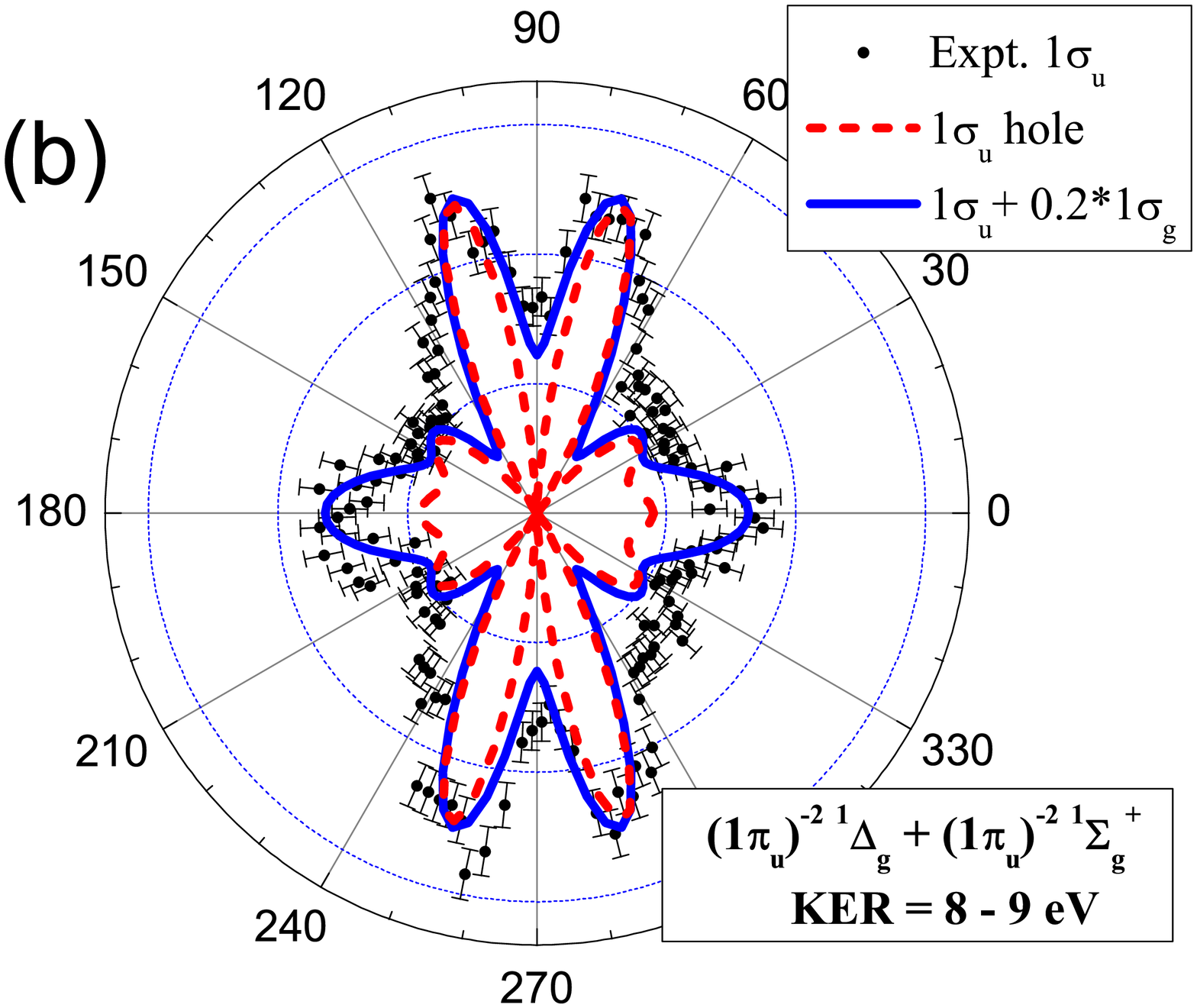}
\caption
{ (Color  online)
 The same as in Figs. 5c,d for KER energies from 8 to 9
eV.  Theoretical  Auger transitions to  the  doubly  charged
molecular  ion  states  $(1\pi_u)^{-2}$ $^1\Delta_g$  
and $(1\pi_u)^{-2}$ $^1\Sigma_g^+$ are included.
}
\end{figure}

     Finally,  at KER between 10.2 and 11 eV  three
terms are contributing to the angular distributions shown in
Fig. 7,  namely, $(2\sigma_u)^{-1} (1\pi_u)^{-1}$ $^1\Pi_g$, 
$^1\Sigma_g^+(3)$ (see Fig.  2), and 
$(3\sigma_g)^{-1} (2\sigma_u)^{-1}$ $^1\Sigma_u$. 
The last term is  responsible  for several   discrete  
transitions  appearing  at   these   KERs.  The  
characteristic  features  of  these  angular distributions  
are  defined basically by the  
$(2\sigma_u)^{-1} (1\pi_u)^{-1}$ $^1\Pi_g$ term. Namely, 
this term gives the main contribution  at the  angles $70^\circ$ 
and $110^\circ$ for the $1\sigma_g$ hole state and at $45^\circ$,
$90^\circ$, and $135^\circ$ for the $1\sigma_u$ hole state. 
As to  the  $^1\Sigma_g^+(3)$ state,  from  the  calculations 
of {\AA}gren [37]  follows  that though the main configuration 
contributing to this state  is $(2\sigma_u)^{-2}$, the admixture 
of other configurations like  $(1\pi_u)^{-2}$ is  substantial. 
Since in our calculations the configuration interaction is not 
taken into account, we included into our fitting  two separate 
configurations, $(2\sigma_u)^{-2}$ and  $(1\pi_u)^{-2}$.
The  ratio  of  different theoretical contributions  to  the
fitted  curve is 
$I(^1\Pi_g) : I(^1\Sigma_g^+) : I(^1\Sigma_u) = 1 : 0.2 : 0.12$. 
The results of fitting are again in a  reasonable agreement 
with the experiment.
\begin{figure}
 \includegraphics[width=7.5cm,angle=0]{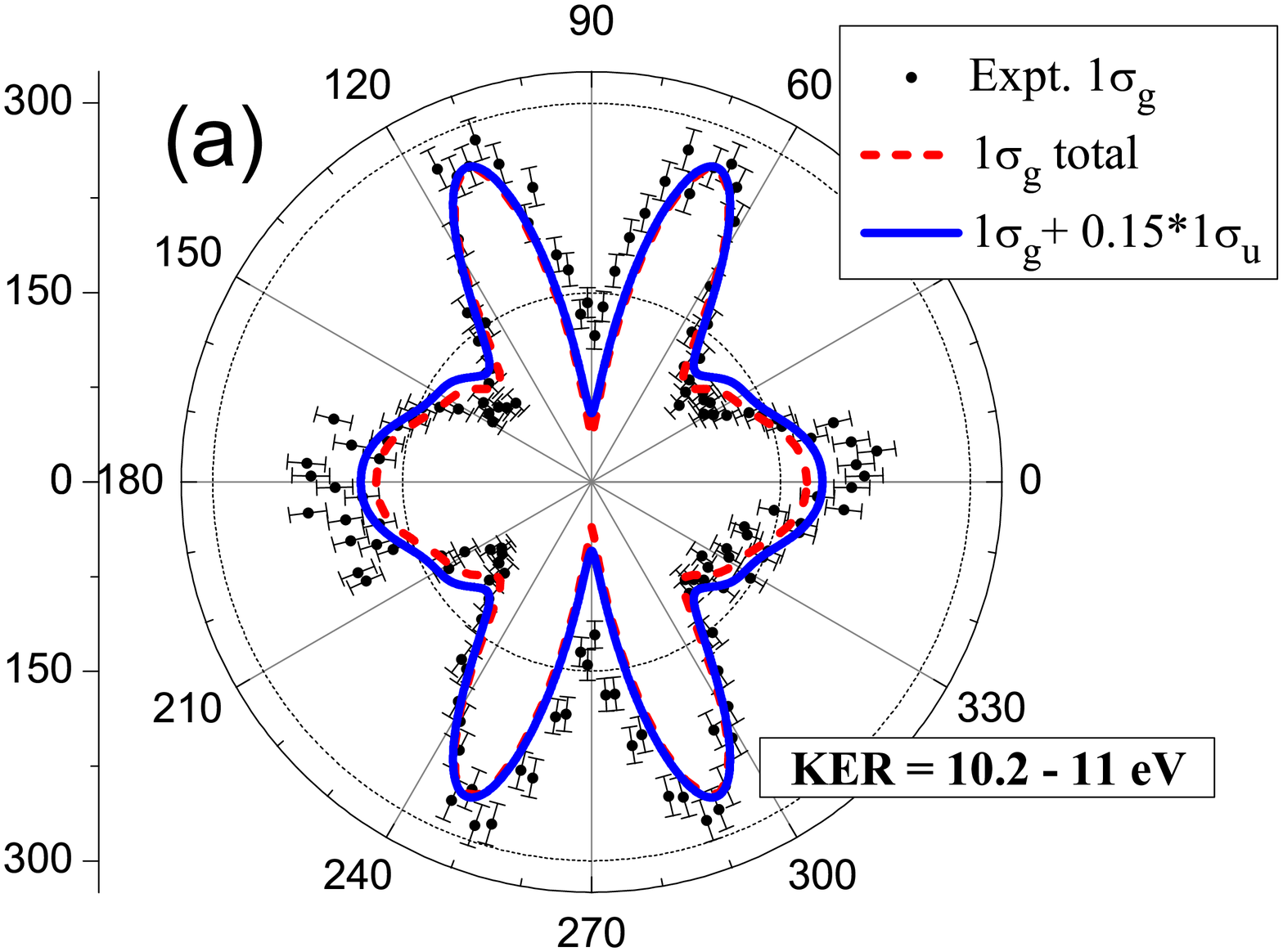}
 \includegraphics[width=7.5cm,angle=0]{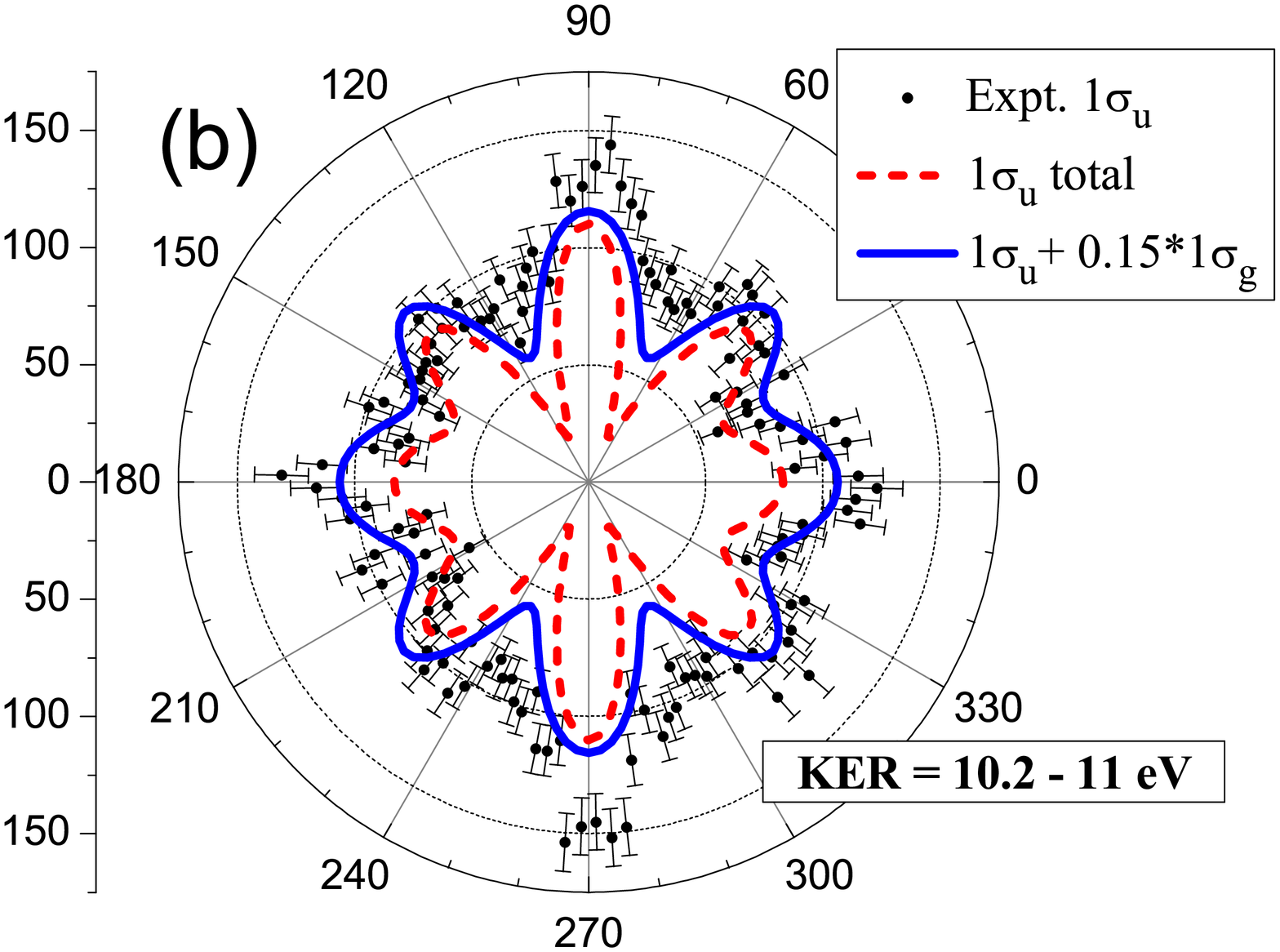}
\caption
{ (Color  online)
 The same as in Figs. 5c,d for KER energies from 10.2  to
11  eV.  Theoretical Auger transitions to the doubly charged
molecular  ion states 
$(2\sigma_u)^{-1} (1\pi_u)^{-1}$ $^1\Pi_g$, 
$(1\pi_u)^{-2}$ $^1\Sigma_g^+$  and
$(3\sigma_g)^{-1} (2\sigma_u)^{-1}$ $^1\Sigma_u$ are included.
}
\end{figure}

     The  contributions  of different triplet  final  states
have  not been identified in our fittings due to their small
contribution  as  was already mentioned  earlier.  Remaining
difference between theory and experiment can be explained by
approximations accepted in our calculations. In  particular,
the   calculated   Auger   electron  angular   distributions
correspond  to  a  fixed equilibrium internuclear  distance,
while  in  experiment the internuclear distance  is  varying
inside  the FC region. Evidently, the Auger electron angular
distributions depend on the internuclear distance.

     Another   source  of  error  is  connected   with   the
description  of  the  doubly charged  final  ion  state.  We
calculated  the  angular distributions for  a  well  defined
configuration  of  the final states, while  calculations  of
potential   energy  curves for  N$_2^{2+}$ demonstrated  that
configuration  mixing  plays  an  important  role   [34-37].
Fortunately,  the  main final states, giving  the  principal
contribution  to the Auger electron spectra,  namely  
$(3\sigma_g)^{-1} (1\pi_u)^{-1}$ $^1\Pi_u$, 
$(1\pi_u)^{-2}$ $^1\Sigma_g^+$, $(1\pi_u)^{-2}$ $^1\Delta_g$, 
and $(2\sigma_u)^{-1} (1\pi_u)^{-1}$ $^1\Pi_g$,  
can  be  represented sufficiently  well  by  a  single
configuration [37].

   When  separating the contribution of the $1\sigma_u$ state  at
the  angles $115-150^\circ$ and $295-330^\circ$ it is evident from 
Fig. 1 that the contribution of the $1\sigma_g$ state is not 
negligible, so that the neglect of the interference term in 
equation (3) is not well justified. But its inclusion makes the 
calculations much  more  laborious.  One  can  mention  also  a  
possible contribution of many electron correlations  beyond the 
HF approximation used in this paper. It is difficult to give  a
numerical estimate of all these effects. Since the degree of
agreement  between theory and experiment  in  figs.  5-7  is
quite  satisfactory,  all possible theoretical  improvements
mentioned above hardly can change the principal conclusions.

\subsection{Discussion of KER spectrum}
      Fig.  8 shows the total KER spectrum for all Auger decay 
channels (that  is  without  coincidence with
photoelectrons and integrated over the angle $\theta$ relative  
to the  molecular axis). This spectrum contains several strong
discrete lines and a continuous contribution. Qualitatively,
this  spectrum  is similar to the KER spectrum  observed  in
[11]  by  electron  scattering.  According  to  the  results
demonstrated above, a broad maximum between 7 and 10 eV is mainly 
formed by the transition to the $(1\pi_u)^{-2}$ $^1\Delta_g$ 
state. It coincides  with the region where the corresponding 
potential energy  curve crosses the Frank-Condon region 
(see Fig. 2). To trace  the contribution of this final state  
more precisely, we selected from the data shown in Fig. 4 the
angles   corresponding  to  the  Auger   electron   emission
perpendicular  to  the molecular  axis ($85-95^\circ$).  The
corresponding results are shown in Figs. 9c,e for the $1\sigma_g$
and $1\sigma_u$ hole states separately, and in Fig. 9a for the  
sum of these two states. As is evident from theoretical angular
distributions shown in  Fig. 3, for the $1\sigma_g$ hole state
practically only  one $^1\Delta_g$ term is contributing in this
direction.  According  to Fig. 9c, this  contribution  as  a
function  of  KER  at  first increases, and  then  decreases
inside the FC region which is in accord with the behavior of
the zero order vibrational wave function of the ground state
of N$_2$.
\begin{figure}
 \includegraphics[width=7.5cm,angle=0]{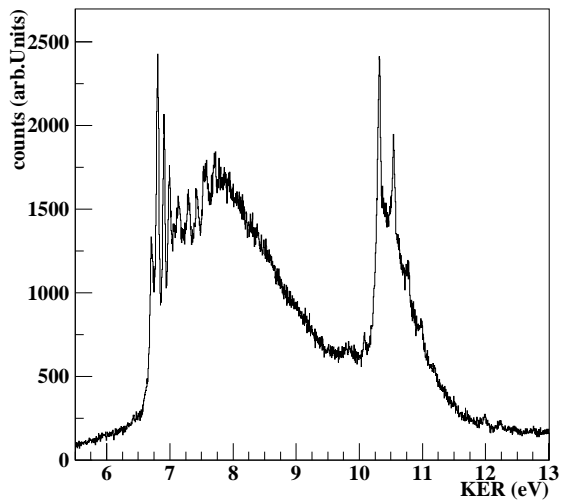}
\caption
{
 Total experimental Auger electron KER spectrum  (in
arbitrary units), which is the sum of contributions  of  the
$1\sigma_g$  and $1\sigma_u$ states  integrated over  all  
Auger  electron emission angles.
}
\end{figure}
\begin{figure}
 \includegraphics[width=7.5cm,angle=0]{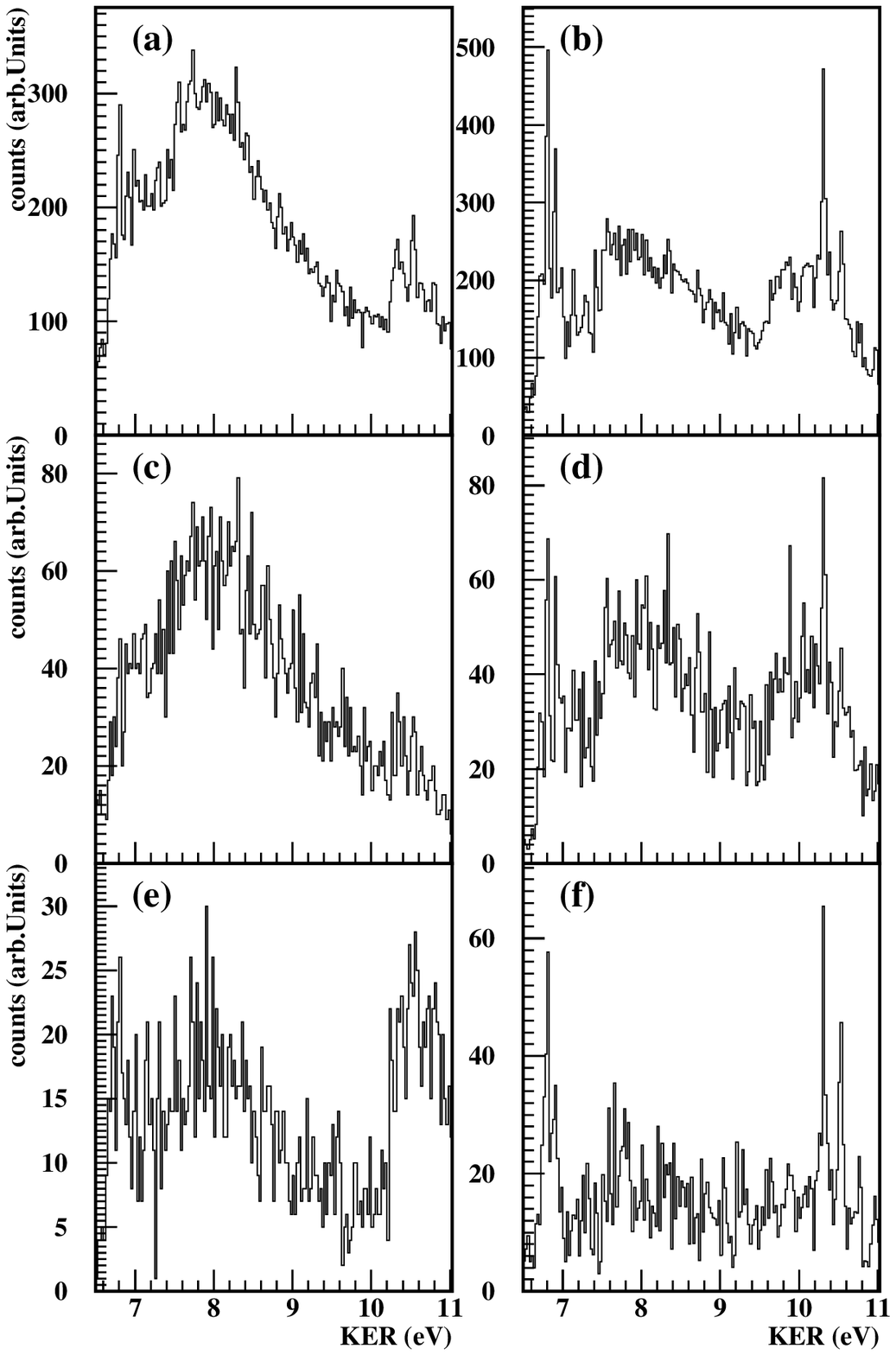}
\caption
{
  Auger  electron intensities  integrated  over  the
angles  $90^\circ \pm 5^\circ$, that is perpendicular to the 
molecular  axis ((a),  (c), and (e), left column), 
and over the angles $0^\circ \pm 5^\circ$ plus 
$180^\circ \pm 5^\circ$, 
that is parallel to the molecular axis  ((b),
(d),  and (f), right column). 
Upper line: without separation of the contributions of the 
$1\sigma_g$ and $1\sigma_u$ hole states ((a) and (b)),  
the second line: contribution of only $1\sigma_g$ hole 
state ((c) and (d)), 
the third line: contribution of only $1\sigma_u$ hole
state ((e) and (f)).
}
\end{figure}

      In  the  KER energy region between 6.8 and  7.5  eV  a
substantial  contribution in Fig. 8 is  given  also  by  the
transition to the $(3\sigma_g)^{-1} (1\pi_u)^{-1}$ $^1\Pi_u$ 
final state. This is in
agreement  with the position of the corresponding  potential
energy curve inside the FC region in Fig. 4. It is important
to  mention  that  due  to  the potential  barrier  (at  the
internuclear distance about 1.8 {\AA}) the contribution  of 
this state  inside the  FC region  is  visible  only  at the
internuclear distances smaller than 1.1 {\AA}. Due to that its
contribution has a sudden jump at KER$\approx$ 6.8 eV, at lower  
KER energies the fast dissociation is not possible. This  sudden
jump is a characteristic feature of the KER spectrum in Fig. 8.  
Finally,  the maximum between 10.3 and 12 eV  is  formed
mainly by the $(2\sigma_u)^{-1} (1\pi_u)^{-1}$ $^1\Pi_g$ state. 
This contribution is also seen in  Fig. 9e where there is a sharp  
increase  of intensity starting from 10.3 eV. This is in accord 
with  the behavior  of the  theoretical  Auger electron  angular
distribution for the $^1\Pi_g$ term for the $1\sigma_g$ hole  
state  which has a maximum at the angle $90^\circ$ (see Fig. 3). 
There are two calculations  of the potential energy curve for  
this  state shown  in  Fig. 2 which do not coincide well within  
the  FC region.  The sharp increase of the Auger electron  
intensity at  KER=10.3 eV definitely testifies to the presence 
of some potential  barrier like in the case of the $^1\Pi_u$ 
final  state,
or  at  least  to a non-monotonic decrease of the  potential
energy  curve with increasing internuclear distance like  in
the  calculations of Taylor [35]. But in the latter case the
position of the potential energy curve inside the FC  region
does not fit the position of the maximum in the experimental
KER  spectrum.  Therefore  we conclude  that  the  potential
energy  curve  for  the $^1\Pi_g$ dicationic  state  needs  
to  be calculated more accurately.

  The contribution of the $(1\pi_u)^{-2}$ $^1\Sigma_g^+$ state  
does not produce  a well separated maximum in the total KER  
spectrum shown in Fig. 8.  To separate  the contribution of 
$\Sigma$ states we show  in Figs. 9b,d,f the parts of the spectrum 
of Fig. 4  in the  direction of the molecular axis, that is in 
the regions $-5^\circ - +5^\circ$ and $175-185^\circ$. For 
the $1\sigma_g$ hole state in Fig.  9d there  are  two broad 
maxima in the Auger electron intensity along the molecular axis 
which must be connected with the $\Sigma$ terms  ($\Pi$  
and $\Delta$ terms do not contribute along the molecular
axis).  The first of these maxima corresponds to the 
$(1\pi_u)^{-2}$ $^1\Sigma_g^+$  state, which is in agreement 
with the behavior of  the corresponding potential energy 
curve shown in  Fig.  2.  The second  maximum  is most 
probably produced  by  the  $^1\Sigma_g^+(3)$ state also shown 
in Fig. 2 which is connected mainly  with the $(2\sigma_u)^{-2}$ 
configuration. Its position in Fig. 2 is shifted
upwards by about 4 eV as compared to other potential  energy
curves  contributing  at KER energies studied  by  us  which
means that the corresponding state dissociates into the pair
of excited atomic ions N$^+(^1 D)$+N$^+(^1 D)$ or into 
N$^+(^3P)$ + N$^+(^1S)$.

     From   the  analysis  of  the  Auger  electron  angular
distributions  presented in Fig. 4 we can  conclude  that  a
strong  discrete transition at KER = 6.8 eV  corresponds  to
the $(2\sigma_u)^{-1} (3\sigma_g)^{-1}$ $^3\Sigma_u$ state. 
Two  other  strong  discrete transitions  at KER=10.32 and 
10.54 eV can be  unambiguously indentified as transitions 
to the $(2\sigma_u)^{-1} (3\sigma_g)^{-1}$ $^1\Sigma_u^+$ state
which  is in agreement with the identification made  earlier
by  Lundqvist et al [11]. This is supported by the  presence
of  contribution of these lines along the molecular axis  in
Figs. 9d,f for both $1\sigma_g$ and $1\sigma_u$ shells in accord  
with  the corresponding  theoretical angular  distributions  
shown  in Fig. 3.

\section{Conclusions}
     We  demonstrated that the measurement in coincidence of
photoelectrons and Auger electrons together with the  singly
charged  atomic  ions  produced by  dissociation  of  doubly
charged molecular ion allows separating Auger decay channels
corresponding to the $1\sigma_g$ and $1\sigma_u$ hole states 
of  N$_2$  without need  to  separate these transitions 
in energy. In addition,
it  becomes  possible  to disentangle the  contributions  of
different repulsive doubly charged molecular ion states as a
function  of  KER  energy by comparison  with  corresponding
theoretical  Auger  electron angular  distributions  in  the
molecule  fixed  frame. This allows to follow experimentally
the  behavior of the potential energy curves for  dicationic
final states within the Frank-Condon region. Presentation of
the  Auger  electron angular distributions as a function  of
kinetic  energy  release  of two atomic  ions  opens  a  new
dimension  in the study of Auger decay. In particular, one 
can follow the contribution of a given Auger transition as 
a function of internuclear distance. To the best  of  our
knowledge,  that  can not be done by any other  method.  The
strongest  discrete  lines can also be  identified  by  this
method.  The method can be used at different photon energies
and  with  different light polarization,  though  circularly
polarized  light  seems  to  give  a  better  resolution  of
contributions  of  the $1\sigma_g$ and $1\sigma_u$ hole  
states.  Evidently, this  method  is  applicable to other  
homonuclear  diatomic molecules.
     
\section{Acknowledgments}
We  acknowledge  outstanding support by  the  staff  of  the
Advanced  Lights Source in particular by Hendrik  Bluhm  and
Tolek  Tyliszczak. The work was supported  by  the  Deutsche
Forschungsgemeinschaft  and by the office  of  Basic  Energy
Sciences, Division of Chemical Sciences of the US DOE  under
contract   DE-AC03-76SF00098  and   DE-FG02-07ER46357.   NAC
acknowledges    the    financial   support    of    Deutsche
Forschungsgemeinschaft through a Mercator professorship. SKS 
and NAC acknowledge the  hospitality  of the Goethe University 
in  Frankfurt  am Main and the financial support of  RFBR
(grant No 09-03-00781-a).

\end{document}